\documentclass[12pt,preprint]{aastex}

\newcommand{\msol}{\mathrm{M_{\odot}}}
\newcommand{\rsol}{\mathrm{R_{\odot}}}
\newcommand{\thco}{^{13}\mathrm{CO}}

\newcommand{\hii}{\mathrm{H\,\scriptstyle II}}
\newcommand{\kms}{{\mathrm{km \, s^{-1}}}}
\newcommand{\de}{{^{\circ}}}
\newcommand{\ls}{\mathrel{\raise0.35ex\hbox{$\scriptstyle <$}\kern-0.6em
\lower0.40ex\hbox{{$\scriptstyle \sim$}}}}
\newcommand{\dv}{|{\Delta v}|}

\slugcomment{For submission to Ap J L }

\shorttitle{GMCs in Spiral Arms}
\shortauthors{Stark and Lee}

\begin{document}

\title{Giant Molecular Clouds are More Concentrated 
to Spiral Arms than Smaller Clouds
}

\author{Antony A. Stark}
\affil{Smithsonian Astrophysical Observatory, Cambridge MA 02138}
\email{aas@cfa.harvard.edu}

\and

\author{Youngung Lee} 
\affil{Korea Astronomy Observatory, Taeduk Radio Astronomy
Observatory, Daejeon, Korea}
\email{yulee@trao.re.kr}

\begin{abstract}
From our catalog of Milky Way molecular clouds,
created using a temperature thresholding algorithm
on the Bell Laboratories $\thco$ Survey, we
have extracted two subsets:
(1) 
Giant Molecular Clouds (GMCs),
clouds that are definitely larger
than $10^5 \, \msol$, even if they are at their ``near distance'',
and
(2) clouds that are definitely smaller than
$10^5 \, \msol$, even if they are at their ``far distance''.
The positions and velocities of these clouds are compared to
the loci of spiral arms in $\ell,\, v$ space.   
The velocity separation of each cloud from the
nearest spiral arm is introduced as a ``concentration statistic''.
Almost all of the GMCs are found near spiral arms.
The density of smaller clouds is enhanced near spiral arms, 
but some clouds ($\sim\,10\%$) are unassociated with any
spiral arm.
The median velocity separation between 
a GMC and the nearest spiral arm is $3.4\pm0.6\, \kms$, 
whereas the median separation between smaller clouds
and the nearest spiral arm is $5.5\pm0.2\, \kms$.

\end{abstract}


\section{Introduction}

The Galaxy as a whole affects star formation
through the mechanism of spiral structure.
Stars in galactic disks
tend to develop spiral-shaped density waves that arise
either
through spontaneous instability 
or through
gravitational perturbations from
nearby galaxies or a central bar.
The spiral density wave modulates the gravitational
potential in the disk, and the interstellar medium reacts 
non-linearly to this varying potential,
gathering and concentrating the gas.
Giant Molecular Clouds (GMCs) form, leading to a local increase in
the star formation rate and the creation of giant $\hii$ regions.
The cloud formation process is not well understood, and 
is the subject of
ongoing investigation \citep{ 
elmegreen00,pringle01,hartmann01,zhang02,elmegreen02,ostriker04}.

In this {\em Letter}, we measure the
degree to which GMCs and smaller molecular clouds
are concentrated in the spiral arms of
the Milky Way, to provide a quantitative comparison with theoretical
models of molecular cloud formation.  
In \S 2, we take the Bell Laboratories $\thco$ Survey data 
\citep{lee01} and use a
brightness temperature thresholding algorithm to create a 
catalog of molecular clouds.
From this catalog, we extract a set of clouds that
are definitely GMCs, and a set of clouds that are definitely
smaller than GMCs.
In \S 3, we use surveys of neutral and ionized hydrogen to
define the spiral arms.
In \S 4, we compare the locations of clouds
in $(\ell,\,v)$-space with the spiral arm loci.
We show that the GMCs and the
smaller clouds have significantly different distributions.
The GMCs are more concentrated to the spiral arms than
the smaller clouds.

\section{Cloud Identification and Mass Estimation}

A catalog of clouds was generated from the
Bell Laboratories $\thco$ survey data  
\citep{lee01} using a 
thresholding technique 
described in \citet{stark05c}.
This survey contains $2.3\,\times\,10^7$ pixels of $\thco$ data,
each $0.05\de \,\times\,0.05\de\,\times\,0.68\,\kms$ in size.
Survey pixels having $T_R^{*} > T_{\mathrm{th}}$
are identified
and grouped together in $(\ell, \, b, \, v)$ 
space to make a ``cloud", where all
the pixels constituting the cloud are above the threshold 
and also adjacent to at least one other pixel that
is also above the threshold.
A ``cloud" is then a connected volume of pixels, 
all of which are above the threshold.
Applying the thresholding method with 
$T_{\mathrm{th}} = 1 \mathrm K $ on the
Bell Laboratories $\thco$ Survey
yields a catalog of 1,400 clouds.
As further described in \citet{stark05c}, we assign several 
possible distances to each cloud based on our knowledge
of the velocity field of the Galaxy.  
Essentially, the distances correspond to the
``near'' and ``far'' points  of 
the distance ambiguity (cf. Mihalas 1968\nocite{mihalas68}), 
plus some
additional uncertainty due to 
random motions.
Each cloud has a range of possible distances.

We want to distinguish GMCs from smaller clouds, but the
distances are uncertain, and so are the luminosities and estimated masses.
We can, however, derive a subset of the catalog that
contains only GMCs, and another subset that that contains
only clouds that are definitely smaller than a GMC.
For the purposes of this {\em Letter}, we define a GMC
as a member of our catalog with
$L(\thco) \,>\, 5 \times 10^3 \, \mathrm{K \, km \, s^{-1} \, pc^2}$.
The mass corresponding to this luminosity is 
$1 \times 10^5\, \mathrm{\msol}$, based on the
mass-luminosity relation 
$M_c = [20 \, \mathrm{\msol / K \, km \, s^{-1} \, pc^2}] L(\thco)$
derived 
in \citet{stark05c}
for clouds identified by the method we use here.
For each cloud in the catalog, we calculate the
range of luminosities corresponding to 
the range of possible distances.
If all these possible luminosities exceed the luminosity
threshold
$L(\thco) \,>\, 5 \times 10^3 \, \mathrm{K \, km \, s^{-1} \, pc^2}$,
the cloud is definitely a
GMC, regardless of the distance uncertainties.  
Applying this criterion to the catalog yields 56 GMCs.
If all the possible luminosities fall below the threshold,
the cloud is definitely not a GMC, and 
the cloud is included in the ``definite small cloud'' set.
This second criterion yields 1257 small clouds.
Our catalog also contains 87 clouds that do not
fulfill either criterion.  These are moderately-large
clouds whose distance is not well determined, and
we exclude them from further consideration.

\section{Spiral Arm Loci}

Spiral arms in the Milky Way have been identified 
in surveys of 21 cm atomic hydrogen and radio
recombination lines.  Here we adopt the analyses of
\citet{reifenstein70}, 
\citet{burton70}, 
\citet{shane72}, 
\citet{lindblad73}, 
and
\citet{simonson76}, to determine the locations
of the spiral arms in 
$(\ell, \, v)$ space:
we take locations defining the spiral arms from
these references, and interpolate using cubic spline functions.
The loci of the arms are well-determined for
$ 20\de \, < \, \ell \, < \, 140\de$, but not
for the Galactic Center Region.
We therefore exclude from further analysis all clouds with
$ \ell < 20\de$.
This reduces the samples to 39 GMCs and 932 smaller clouds.
The locations of the clouds, and the locations of the
spiral arms, are plotted in figure 1.
The spiral arms are identified by letters, as in \citet{cohen80}.

Drawing the spiral arms onto the $(\ell, \, v)$ 
diagram of the Milky Way has been controversial.  The criteria adopted 
by the above authors for the placement of the arms
are subjective and ill-defined.  Arms ``A'' and
``B'', the local arm and the Lindblad Ring,
are not large-scale features of the Galaxy, but local
spurs that loom large because they are close to the Sun.
Spiral arms ``A'' through ``F'' are adopted here because
they are traditional and because there is no alternative; the analysis
below will, however, be seen to support this initial choice.

Almost all the GMCs (red circles) lie close to
spiral arms.  The most notable exceptions are the GMCs in
the bridge between the Sagittarius (C) and Scutum (D)
arms near $\ell = 38\de$, $v = 90\,\kms$.  The local
spiral arm and Lindblad Ring (A and B) contain no GMCs within
the area of sky covered by the Bell Labs survey.  
As noted by \citet{cohen80},
the Perseus arm (F) is
particularly distinct and well-separated from surrounding
material.

\section{Concentration of Clouds to the Arms}

There is at least one spiral arm at some velocity 
for each value of $\ell$ in figure 1.
We can therefore define a ``concentration statistic'' for each
cloud:  the absolute value of the separation in velocity 
between that cloud and the nearest spiral arm, $\dv$.
The set of $\dv$ for each of our cloud samples is the
distribution of separations of those clouds 
from the spiral arm loci. These distributions are plotted
in figure 2 for the GMC and small cloud samples.
The small cloud distribution has a 
long tail, containing about 10\% of all clouds, 
that extends  to large values of $\dv$.
Most of these
are near $\ell \,\approx \,55\de, \,v\,\approx\,30\,\kms$,
where we have a long line of sight that falls between
spiral arms and includes the tangent velocity, where the
phase space density is high.
These are clear examples of interarm clouds.
All but five of the GMCs have $\dv \, < \, 14 \, \kms$.
Of those five, four are clouds
in the Sagittarius-Scutum bridge \citep{cohen80}.

The Kolmogorov-Smirnov statistic \citep{press92} is the largest 
vertical separation
between the two distributions, as shown in figure 2.
The hypothesis that the two sets of $\dv$ are drawn
from the same parent distribution is
rejected at the 94\% level by the Kolmogorov-Smirnov 
test --- it is highly likely that the GMCs
are distributed differently with respect to the spiral arms
than are the small clouds.
The two distributions have significantly different medians.
The median $\dv$ for the
GMCs is $3.4\pm0.6\, \kms$, 
whereas the median $\dv$ for the small clouds
is $5.5\pm0.2\, \kms$.  The errors in the medians are estimated
by the bootstrap method \citep{efron83}.

We will compare these results with
the concentration statistics of
objects that have the same radial distribution and
velocity dispersion as
molecular clouds but
are azimuthally symmetric, and therefore have no
concentration to spiral arms.
The radial distribution can be parameterized
by:
\begin{equation}
\rho \, \propto \, R^\alpha \, 
\mathrm{exp}\left(-{{R}\over{R_\mathrm{s}}}\right) \, ,
\end{equation}
where $\rho$ is the surface density of objects, $R$ is radius from
the galactic center, and $\alpha$ and $R_\mathrm{s}$ are
parameters.
The galactic rotation curve is taken to be flat, at a 
velocity $\Theta_0$.
The velocity dispersion is Gaussian, with a
one-dimensional root-mean-square of $\sigma_v$ \citep[e.g.,][]{stark89b}.
A set of Monte Carlo objects were generated by
the following steps:
(1) Choose a set of parameters, $\alpha$, $R_\mathrm{s}$, $\Theta_0$
and $\sigma_v$.
(2) Use a random number generator 
to create a set of objects in two dimensions that have
a radial distribution given by equation 1. 
(3) Calculate the $\ell$ and $v$ of each 
object as seen from the Sun at $\rsol = 8\,\mathrm{kpc}$.
(4) Generate a random Gaussian deviate for each
object from a distribution
with $\sigma \, = \, \sigma_v$ and zero mean, and add it to each $v$.
(5) Compare the resulting set of random $(\ell, v)$ 
values to the catalog of actual molecular clouds, using
a two-dimensional Kolmogorov-Smirnov test \citep{press92}.
(6) Go back to step 1, varying parameters to maximize the
similarity between the random set of objects and
the observed catalog.

This procedure results in the parameters 
$\Theta_0 \, = \, 215 \, \kms$,
$R_\mathrm{s} \, = \, 1.7 \, \mathrm{kpc}$, $\alpha = 2.3$,
and $\sigma_v \,=\, 7.7\,\kms$.  
This value of $\Theta_0$ should not be
taken to be a measure of the Sun's rotational velocity.  If
the rotation curve were parameterized as
$\Theta(R) \,\approx\, \Theta_0 \,+\, (R/\rsol)\Theta_1 \,+\,\ldots\,$,
then 
$\Theta_1$ cannot be constrained by fitting to objects
in the inner Galaxy: galactic kinematics looks the same
with a solid-body term added to the rotation
curve.
These data do not constrain $\alpha$ very well either, since
our cutoff at $\ell\, = \, 20\de $ eliminates much of the molecular
hole at $R\,\approx\,3.5\,\mathrm{kpc}$.  The shape of the
inner edge of this hole is parameterized by $\alpha$, and
almost all values of $\alpha$ between 2 and 3 fit well.
The quick drop-off of the molecular ring to larger radii, 
$R_\mathrm{s} \, = \, 1.7 \, \mathrm{kpc}$, is typical of
fits to first quadrant molecular line surveys
\citep{burton78,solomon79}.
The value of 
$\sigma_v \,=\, 7.7\,\kms$ is consistent with the value  
$\sigma_v \,=\, 7.8^{+0.6}_{-0.5}\,\kms$ determined for 
moderately-large
clouds near the Sun by \citet{stark89b}.

The distribution of $\dv$ values for the Monte Carlo objects
is plotted in green in figure 2, and their median value of $|\Delta v|$
is $7.81\pm0.05\,\kms$.  This value is robust, in the sense that essentially
any azimuthally-symmetric distribution of random objects 
in the galactic plane that we have tried will produce
a median value of $|\Delta v|$  that is greater than $7\,\kms$.
The small molecular clouds are more concentrated to spiral arms than
random objects, and the GMCs are even more concentrated than the
small clouds. 

\section{Discussion} 

It is unfortunate that we cannot observe the Milky Way from the
outside, and see the spatial relationship between
the molecular clouds and the spiral arms directly. 
Interferometric observations of CO in other galaxies
having well-defined spiral structure
shows that the CO emission is
concentrated to the spiral arms \citep{murgia05}.
In the Milky Way, most of the CO emission is from GMCs.
This is illustrated in figure 3.  Here
we have taken the catalog subset discussed in \citet{stark05c},
where the clouds were selected to have well-determined distances,
and further restricted that subset to clouds with $\ell\,>\,20\de$.
Figure 3 shows the distribution of virial masses for these clouds.
The vast majority of the CO-emitting material, over 85\%, is in the GMCs.
Since the CO luminosity per molecule is approximately constant
\citep{liszt84}, the majority of the CO luminosity will
also arise in GMCs.
It seems likely that this is true of all spiral galaxies, and
that the concentration of CO to spiral arms is
principally a concentration of GMCs to the spiral arms. 

Given the median velocity separations derived in \S4, we
can estimate the typical size of the spatial deviations
from the spiral arms.
The distributions of $\Delta v$ values are approximately
Gaussian for small values, ignoring the long tails.
For purposes of this estimate, we assume that
the spatial deviations from the arms are also Gaussian.
Let $\sigma_a^2$ be the Gaussian velocity variance; its 
square root is proportional
to the observed median, $m$:
\begin{equation}
2\int_0^m\,{{1}\over{\sigma_a\,\sqrt{2\pi}}}\,\mathrm{e}^{-{{v^2}\over
{2\,\sigma_a^2}}}\,\mathrm{d}v \,=\, 
\mathrm{erf}\left({{m}\over{\sqrt{2}\,\sigma_a}}\right)
\,=\,{{1}\over{2}} \,,
\end{equation}
or $\sigma_a = m/(b\sqrt2) = 1.48 \, m$, where
$\mathrm{erf}(b)\,\equiv\,{{1}\over{2}}$.  So
$\sigma_a\,\approx\,8.2\,\kms$ for the small clouds and
$\sigma_a\,\approx\,5.0\,\kms$ for the GMCs.  
We know that the one-dimensional root-mean-square velocity
dispersions of clouds are $\sigma_v \, \approx \, 7.8 \,
\kms$ for moderate-sized clouds ($M\,\ls\,10^{5}\,\msol$, 
Stark \& Brand 1989) \nocite{stark89b}
and
$\sigma_v \, \approx \, 4 \, \kms$ for GMCs ($M\,>\,10^{5.5}\,\msol$,
Stark \& Lee 2005).  \nocite{stark05c}
This dispersion will, approximately, add in quadrature with
the amount of velocity separation caused by the spatial
deviations and the rotation
curve, $\sigma_r$:
\begin{equation}
\sigma_a^2 \,\approx\, \sigma_v^2 + \sigma_r^2 \, .
\end{equation}
A typical spatial deviation of the clouds from the spiral arm is
then
\begin{equation}
\Delta r \,\sim\, {{\sigma_r}
\over{|{{\mathrm{d}v}\over{\mathrm{d}r}}|}} \, \approx \, 
{{\sqrt{\sigma_a^2 - \sigma_v^2}}
\over{|{{\mathrm{d}v}\over{\mathrm{d}r}}|}} \,\sim \,50 \, \mathrm{pc} ,
\end{equation}
where
${|{{\mathrm{d}v}\over{\mathrm{d}r}}|}\,\sim\,
50\,\kms\,\mathrm{kpc}^{-1}$
is a typical change in radial velocity with distance due to galactic
rotation.  The estimated separation,
$\Delta r \,\sim\, 50\, \mathrm{pc}$, which applies to both the small 
clouds and
the GMCs, is small compared to the scale of a spiral arm. 
We do not want to make too much of this rough estimate, but 
we can say that
aside from the long tail in the
distribution of $\Delta v$ values, 
the data are consistent with 
the hypothesis that the spatial concentration of most
molecular clouds to the spiral arms is high.

The complete picture of the interaction of molecular clouds
with spiral arms is a dynamical process in six-dimensional
phase space.  The Bell Laboratories $\thco$ Survey provides
statistical information about two spatial and one velocity
dimension.  We have seen in \citet{stark05c} that, on average, 
the GMCs are tightly concentrated to the galactic plane,
with a FWHM scaleheight about 30 pc.  We now see that
these same GMCs are concentrated to the spiral arms 
as well. Most clouds are associated
with spiral arms, aside  from the 
outliers that are examples of interarm clouds. 
Among clouds near the arms, the GMCs are significantly more
concentrated to the arms in velocity than are the small clouds.

\acknowledgments

We thank the members of
the Bell Laboratories Radio Physics Research Group during the 
ten-year period of the Bell Labs $\thco$ Survey: R. W. Wilson, 
J. Bally, D. Mumma, W. Bent, W. Langer, G. R. Knapp,
and M. Pound.  
We thank N. Tothill for comments.
This work was supported by Basic Research Program
R01-2003-000-10513-0 of KOSEF, Republic of Korea,  
and 
by the William Rollins Endowment Fund 
of the Smithsonian Institution.

\clearpage

\begin{figure}
\plotone{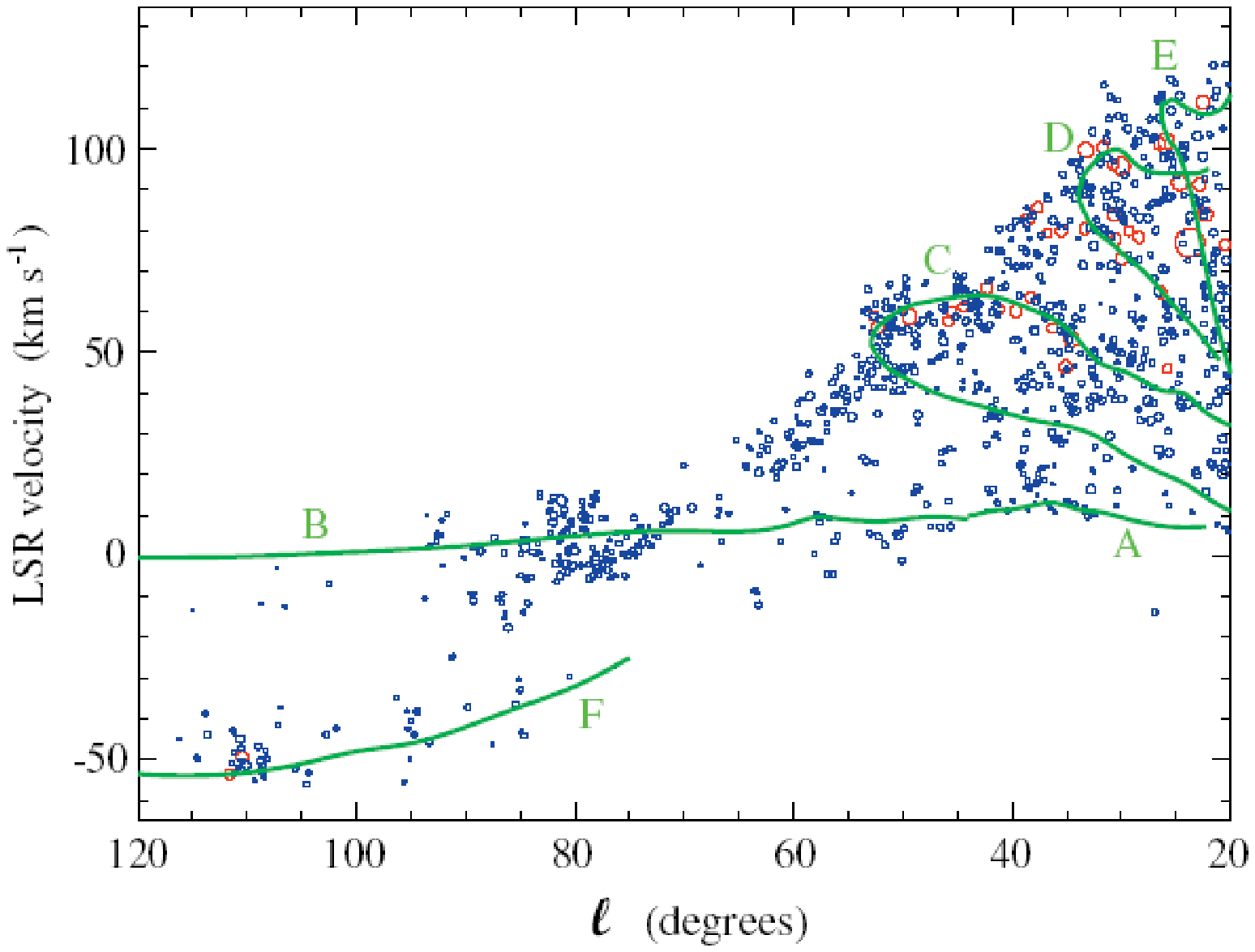}
\caption{Distribution in $\ell$ and $v$ of two sets
of clouds selected from
the Bell Laboratories $\thco$ Survey.
The area of each symbol is proportional to the
velocity width of the corresponding cloud.
The red clouds have a mass  $>\,10^5\,\msol$,
even if they are at their near distance; the
blue clouds have a mass 
$<\,10^5\,\msol$, even if they are at their far distance.
The green curves indicate the loci of spiral arms:
A and B are the local arm, C is the Sagittarius arm,
D is the Scutum arm, E is the 3 kpc arm, and F is the Perseus
arm.
\label{fig1}}
\end{figure}

\clearpage 

\begin{figure}
\plotone{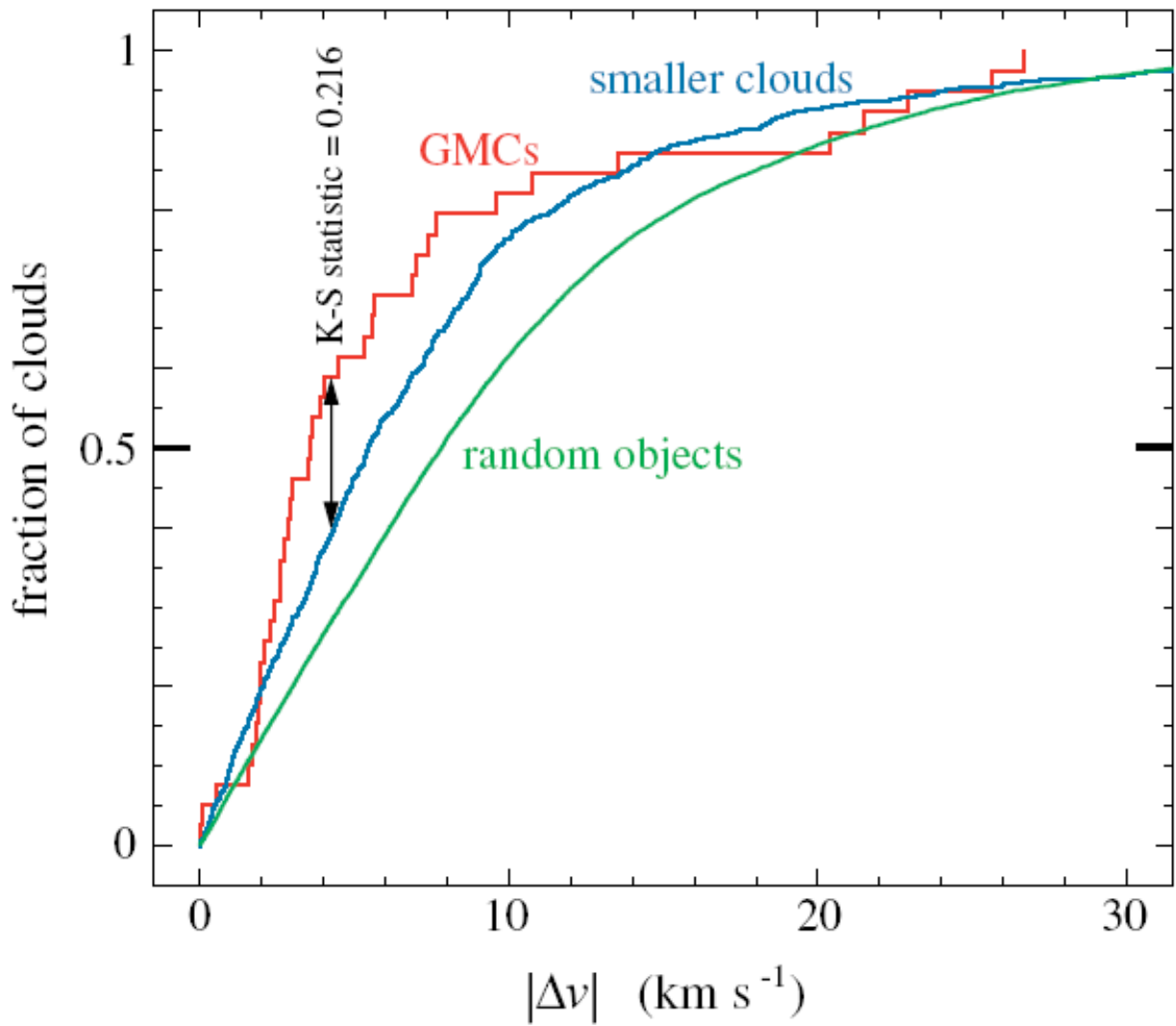}
\caption{Cumulative distribution of separation velocities from the
nearest spiral arm.
For each cloud in figure 1, the velocity separation, $|\Delta v|$,
from the
nearest spiral arm is determined.   The two distributions of these
velocities for 39 GMCs and for 932 smaller clouds are plotted here.
The ordinate is the
fraction of clouds of each size having a separation
velocity less than the value on the abscissa.
The value of the Kolmogorov-Smirnov statistic is shown.
The green line is the distribution of separation velocities
for a set of  azimuthally-symmetric random objects that have the
same radial distribution as the molecular clouds.
\label{fig2}}
\end{figure}

\clearpage 

\begin{figure}
\plotone{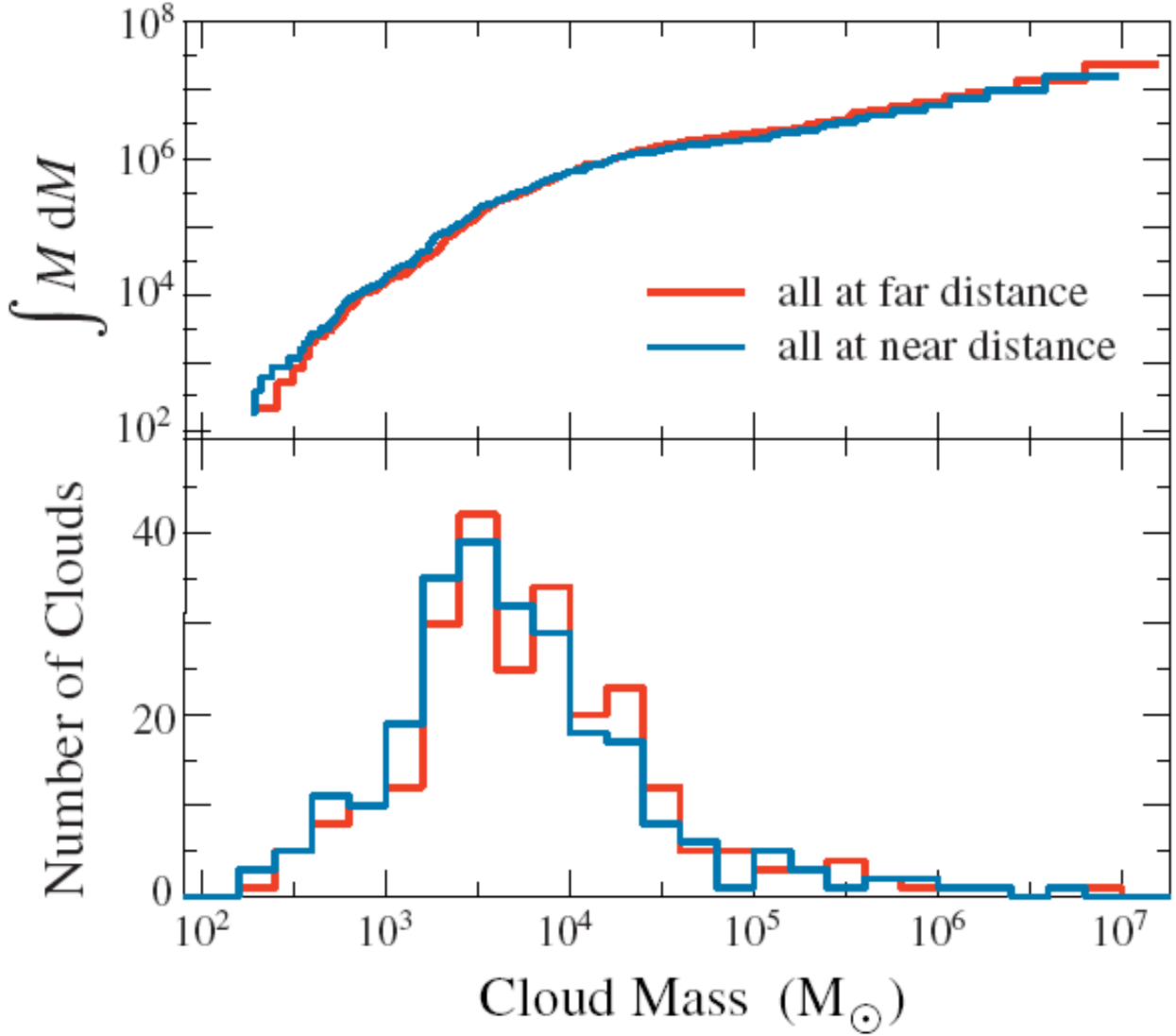}
\caption{
Distribution of estimated virial masses of a cloud sample from the 
Bell Laboratories $\thco$ Survey.  This subset of
clouds was selected to have small distance uncertainty 
in \citet{stark05c}.  The red curves show the distribution
if all the clouds are at their far distance, and the blue
curves show the distribution if all the clouds are at their
near distance.
The lower panel shows a histogram of
the cloud mass distribution in logarithmic bins.  The
coarse sampling grid of the survey may cause it to
miss some clouds smaller than $\sim \, 10^3 \, \msol$.
The upper panel shows the integrated mass distribution.  
Over 85\% of the total mass is in GMCs ($M \, > \, 10^5 \, \msol$).
\label{fig3}}
\end{figure}

\end{document}